\documentclass[a4paper,12pt]{article}
\usepackage{amssymb,latexsym,amsfonts,amstext,amsmath,verbatim}

\font\fiverm=cmr5

\usepackage{prepicte,pictex,postpict}
\usepackage{graphicx}

\newcommand{\frR}{\mathfrak{R}}
\newcommand{\frH}{\mathfrak{H}}
\newcommand{\frK}{\mathfrak{K}}

\newcommand{\bR}{\mathbb{R}}
\newcommand{\bQ}{\mathbb{Q}}

\newcommand{\rmd}{\mathrm{d}}

\newcommand{\rmi}{\mathrm{i}}

\newcommand{\iI}{I^{(\sf o)}}
\newcommand{\iII}{I^{(\sf a)}}

\newcommand{\rmI}{({\sf o})}
\newcommand{\rmII}{({\sf a})}

\newcommand{\AI}{A^{\rmI}}
\newcommand{\AII}{A^{\rmII}}

\newcommand{\phI}{\phi^{\rmI}}
\newcommand{\phII}{\phi^{\rmII}}

\newcommand{\vphI}{\varphi^{\rmI}}
\newcommand{\vphII}{\varphi^{\rmII}}

\newcommand{\frHI}{\frH^{\rmI}}
\newcommand{\frHII}{\frH^{\rmII}}

\newcommand{\xII}{\xi^{\rmII}}

\newcommand{\WI}{W^{\rmI}}
\newcommand{\WII}{W^{\rmII}}

\newcommand{\Ph}{\hat{P}}
\newcommand{\Qh}{\hat{Q}}
 
\newcommand{\sfM}{\sf{M}}

\newcommand{\cA}{\mathcal{A}}

\newcommand{\cM}{\mathcal{M}}
\newcommand{\cO}{\mathcal{O}}

\newcommand{\Ho}{H^{(\sf o)}}
\newcommand{\Ha}{H^{(\sf a)}}
\newcommand{\Hs}{H_{\sf s}}

\newcommand{\nitem}{\vspace{-.4em}\item}

\newcounter{thm}

\newtheorem{thm}{Theorem}[section]
\newtheorem{hypo}[thm]{Hypothesis}
\newtheorem{defn}[thm]{Definition}

\newtheorem{conc}[thm]{Conclusion}

\newtheorem{cond}[thm]{Condition}

\title{Homer nodded once more.  Von Neumann's misreading of the
Compton-Simon experiment and its fallout\thanks{The title of this
paper is inspired by that of Mermin and Schack's paper \cite{MS2018}.
Footnote 17 of \cite{MS2018} reads: `\,``Homer nods''.  \emph{Even the
best of us sometimes slip up}.  From John Dryden's translation of line
359 of Horace's \emph{Ars Poetica: indignor quandoque bonus dormitat
Homerus}.' For more details, see the first paragraph of the article
`Nodding Homer', by William Safire \cite{WS2006}.}}

\author{R. N. Sen\\[1mm]
Department of Mathematics\\ 
Ben-Gurion University of the Negev\\
Beer Sheva 8410501, Israel}
\date{15 February 2023}

\begin{document}

\maketitle

\thispagestyle{empty}

\quad\quad

\vfill\pagebreak

{\small\tableofcontents}
\thispagestyle{empty}
\pagebreak

\section*{\quad}

\addcontentsline{toc}{section}{Abstract}

{
\begin{abstract}

In his book \emph{Mathematical Foundations of Quantum Mechanics}, von
Neumann asserted the following: the Compton-Simon experiment showed
that the state vector must collapse upon measurement of \emph{any}
self-adjoint operator. Comparing von Neumann's account with the
Compton-Simon paper, we find that von Neumann had misinterpreted the
experiment as consisting of \emph{two successive measurements} (which
gave identical results), whereas the experiment only measured two
angles on the same photographic plate.  Note, however, that the state
vector must collapse upon measurement of an \emph{additively conserved
quantity}; otherwise the conservation law could be violated.  Next, it
turns out that the mathematical problem of explaining collapse is not
fully defined until one specifies the nature of the apparatus. If the
apparatus does not have a `classical description', the problem is
insoluble for all observables, even if the measurement is only
approximate (Fine, Shimony, Brown, Shimony and Busch); but if it does,
the problem is soluble \emph{within Schr\"odinger dynamics} (with a
time-dependent hamiltonian) for \emph{additively-conserved
observables}.  The solution, a modification of Sewell's, shows that
the state vector has collapsed, but it \emph{does not reveal} the
eigenvalue of the collapsed state.  The collapse is irreversible, and
results from the interplay of additive conservation laws with the
quantum measurement postulate.  Indeed, the quantum measurement
problem -- as expounded by Wigner -- may be better understood as the
problem of establishing that the two are compatible with each other;
it has little relevance to actual measurements.

\end{abstract}}\vfill

\pagebreak

\numberwithin{equation}{section}

\section{Prologue: The Light is Dark Enough}\label{INTRO}

In his book \emph{Mathematische Grundlagen der Quantenmechanik}
(published 1932) \cite{VN1932},\footnote{In the following, we shall
often refer to this book as \emph{von Neumann's book.}} von Neumann
made an assertion that has spooked the world of physics ever since:
the state vector evolves with time in \emph{two different ways}; when
left alone, it changes smoothly and reversibly, according to the
Schr\"odinger equation with a time-independent hamiltonian; but, upon
measurement of an observable, it \emph{collapses}, abruptly and
irreversibly, into an eigenvector of the observable. Worse, the
collapse happens only when the observation reaches the observer's
brain!\footnote{Von Neumann came to this conclusion because the state
vector of his object-plus-apparatus `combined system' failed to
collapse in his mathematical theory.} 

Wigner was among the few physicists who accepted the observer's
`conscious ego' part in von Neumann's collapse hypothesis.  In an
address to philosophers at a conference in 1961, he said
\cite{EPW1964}:

\begin{quote}

`\ldots physicists have found it impossible to give a satisfactory
description of atomic phenomena without reference to the
consciousness. This\ldots refers to\ldots the process called the
`reduction of the wave packet'. This [reduction of the wave packet]
takes place whenever the result of an observation enters the
consciousness of the observer\ldots'

\end{quote}

\noindent In the next paragraph he added, by way of clarification:

\begin{quote}

`\ldots The interaction between the measuring apparatus and the\ldots{}
\emph{object} of the measurement\ldots results in a state in which
there is a strong statistical correlation between the state of the
apparatus and the state of the object. In general, \emph{neither
apparatus nor object is in a state which has a classical
description} [emphasis added]\ldots'

\end{quote}

Attempts to do away with the observer's conscious ego led to a
variety of \emph{interpretations} of quantum mechanics, none of which
was widely accepted. In the 1960s, the quantum theory of systems with
infinitely many degrees of freedom was formulated
algebraically.\footnote{An accessible summary of the algebraic
formulation, with references, may be found in chapter 2 of Sewell's
book \cite{GLS2002}.} In 1972, Hepp attempted to explain the collapse
of the state vector within this framework \cite{HEPP1972}, using the
concept of \emph{observables at infinity} introduced by Lanford and
Ruelle \cite{LR1969}.  However, in his theory the state vector
collapsed only in the limit $t \rightarrow\infty$, and this fact
allowed Bell to construct a counterexample with an explicitly
time-dependent interaction hamiltonian \cite{BELL1975}.  His article
closes with the statement

\begin{quote}
`\ldots so long as the wave packet reduction is an essential component,
and so long as we do not know exactly when and how it takes over from
the Schr\"odinger equation, we do not have an exact and unambiguous
formulation of our most fundamental physical theory.'
\end{quote}

Forty years later Steven Weinberg (who has a chapter on
`Interpretations of quantum mechanics' in his \emph{Lectures} of 2015)
suggested that the problem may lie with quantum mechanics itself
(\cite{WEINBERG2015}, pp 86--102):

\begin{quote}
`My own conclusion is that today there is no interpretation of quantum
mechanics that does not have serious flaws. This view is not
universally shared. Indeed, many physicists are satisfied with their
own interpretation of quantum mechanics. But different physicists are
satisfied with different interpretations. In my view, we ought to take
seriously the possibility of finding some more satisfactory other
theory, to which quantum mechanics is only a good approximation.'
\end{quote}

A few years later, Sean Carroll speculated that the failure to
`understand quantum mechanics' could impede the resolution of a major
problem in physics: in a \emph{New York Times} article (7 September
2019) with the title and subtitle ``Even physicists don't understand
quantum mechanics.  Worse, they don't seem to want to understand it'',
he wrote \cite{CARROLL2019}:\footnote{The subtitle was clearly a swipe
at those who chose, supposedly, to `shut up and calculate'.}

\begin{quote}

`Gravity, in particular, doesn't fit into the framework of quantum
mechanics like our other theories do. It's possible -- maybe even
perfectly reasonable -- to imagine that our inability to understand
quantum mechanics itself is standing in the way.'

\end{quote}

The remarks quoted above may be summarized as follows. Wigner (1961)
clarified von Neumann's position; Bell (1972) pointed out an
inconvenient truth; Weinberg (2015) suggested that quantum mechanics
could well be a flawed theory, and Carroll (2019) suggested that these
flaws could be responsible for the failure to quantize gravity!
However, in von Neumann's formulation \emph{the collapse hypothesis is
independent of the other principles of quantum mechanics} (as laid
down by him), and it may be accepted, modified or even abandoned
\emph{without altering the basic structure of the latter}.\footnote
{In von Neumann's formulation, \emph{time does not enter into the
construction of the Hilbert space} $\frH = L^2(\bR^n, \rmd^nx)$ (see
pp~197--198 of \cite{VN1932}). He regards it as an external parameter
and denotes the time-dependence of wave functions with a subscript:
$\phi_t$. Operators and commutation relations are defined on $\frH$,
and Born's probability interpretation (the \emph{quantum measurement
postulate} is a part of it) is accepted (p 198 of \cite{VN1932}).
Temporal evolutions are called `interventions' (perhaps because time
is external to Hilbert space), and he introduces \emph{two types}:
(i)~the `2nd', a deterministic, unitary evolution under the
Schr\"odinger equation (apparently with a time-independent
hamiltonian); and (ii)~the `1st', the sudden, probabilistic change, or
collapse, due to a measurement (p 351 of \cite{VN1932}).} That being
the case, it becomes pertinent to ask what led von Neumann to his
collapse hypothesis, and take up the discussion from there.  It should
be stressed that by `quantum mechanics' we mean \emph{nonrelativistic
quantum mechanics}, in which space and time are absolute and there is
no limit on velocities.

In his book, von Neumann interpreted the results of a crucial
experiment by Compton and Simon \cite{CS1925} to mean that the state
vector had to collapse upon measurement of \emph{any} self-adjoint
operator. However, study of the Compton-Simon paper and of von
Neumann's book suggests that (i)~this inference \emph{resulted from a
misreading of the Compton-Simon paper} by von Neumann; (ii)~\emph{the
state vector does collapse upon measurement of an additively conserved
quantity}, but (iii)~this is a consequence of conservation laws, and
\emph{cannot be inferred from the Compton-Simon experiment}. 

{\footnotesize 
\begin{quote} 
{\bf Note.} We adjoin excerpts from the Compton-Simon paper and from
von Neumann's book -- on which our conclusions in
{Sec.}~\ref{CS-VN-INFER} are based -- in Appendix~\ref{SEC-READINGS}.
For ease of reference in what follows, the excerpts from the
Compton-Simon paper are numbered {\bf I}, {\bf II} and {\bf III}.
Those from von Neumann's book -- fifteen consecutive paragraphs -- are
numbered {\bf 1} to {\bf 15}. (Only {\bf 1--9} and {\bf 15} are
reproduced in Appendix \ref{R-VN}, for reasons explained there.)
\end{quote}
}

The rest of this paper is organized as follows.  In
Sec.~\ref{CS-VN-INFER} we describe the Compton-Simon experiment, what
von Neumann inferred from it, and the discrepancies between the
Compton-Simon paper and von Neumann's reading of it. In
Sec.~\ref{COLLAPSE} we discuss which measurements cause the state
vector to collapse. In Sec.~\ref{QMP} we analyse the `quantum
measurement problem', which was declared insoluble in the 1970's and
1980's by Fine, Shimony, Busch and Brown, and was solved by Sewell in
2005. Both assertions were correct, because the authors were dealing
with mathematically distinct problems! It turns out that, using only
Schr\"odinger dynamics, the problem is insoluble (subject to
reservations spelled out in the last paragraph of
Sec.~\ref{SEC-AY-CONC}) if the observable $O$ being measured is an
arbitrary self-adjoint operator and the apparatus is not macroscopic.
If, on the other hand, $O$ is additively conserved and the apparatus
is macroscopic,\footnote{The term \emph{macroscopic} will be defined
more precisely in Sections \ref{APPARATUS} and \ref{SEWELL}.} then the
problem has a solution. The solution only shows that the measurement
causes the state vector to collapse; it does not reveal the result of
an individual measurement.  In Sec.~\ref{SUMMING-UP}, we summarize our
results, and in Sec.~\ref{DISTURBING} we present our general
conclusions, resurrecting a seldom-noticed remark by Einstein and
suggesting a non-reductionist view of the universe which runs counter
to the apparently reductionist view of Wigner, Weinberg and Carroll.
The readings from the Compton-Simon paper and von Neumann's book are
given in Appendix~\ref{SEC-READINGS}.  Bohr's remarkably consistent
views on the subjects concerned are reported in
Appendix~\ref{BOHR-VIEWS}.  Finally, a conjecture about von Neumann's
misreading, some strange misprints in the Beyer translation of von
Neumann's book, some unanswered questions and two private
communications are presented in Appendix \ref{UQ}; they may be of
interest to the historian of physics.

%%%%%%%%%%%%%%%%%%%%%%%%%%%%%%%%%%%%%%%%%%%%%%%%%%%%%%%%%%%%%%%%%%%
%%%%%%%%%%%%%%%%%%%%%%%%%%%%%%%%%%%%%%%%%%%%%%%%%%%%%%%%%%%%%%%%%

\section{The Compton-Simon experiment and von Neumann's
reading of it}\label{CS-VN-INFER}

\subsection{The Compton-Simon experiment}\label{CS-EXPT}

In 1924, before the advent of quantum mechanics, Bohr, Kramers and
Slater (hereafter BKS) published a paper on `The quantum theory of
radiation', in which they tried to reconcile the wave and particle
aspects of radiation by proposing that energy and momentum were
conserved statistically, but not in individual collisions of electrons
with radiation.  Compton and Simon proceeded to test this hypothesis
by an experiment (see footnote \ref{SWANN}), and reported the results
in \cite{CS1925}. The idea of the experiment, described below, is
explained in quotation {\bf I} of Appendix \ref{R-CS}, and the
experimental arrangement, shown in {Fig.}~1, is adapted from {Fig.}~1
of \cite{CS1925}.

In the experiment, an $x$-ray photon of wavelength $\lambda$ enters a
cloud chamber at {\sf A} and is scattered by an electron at {\sf O}.
The scattered photon is scattered again by a second electron (called
the $\beta$-electron by Compton and Simon) at $\sf P$. The cloud
chamber photograph shows the tracks of the recoil and the
$\beta$-electron, which are not straight lines due to repeated
scatterings.  The heavy arrows {\sf OR {\rm and} PB} show the initial
directions of flight of the recoil and the $\beta$-electron
respectively.  The line through {\sf A {\rm and} O} is known from the
geometry of the apparatus, and the line {\sf OP} -- the line of flight
of the scattered photon -- is inferred from the photograph.  Thus the
angles $\theta$ and $\phi$ can be determined from the photograph.

\begin{comment}
\begin{figure}[ht]%
\centering
\includegraphics[width=\textwidth,keepaspectratio]{HOMER-FIGURE.eps}
\caption{Illustrating the Compton-Simon experiment}\label{FIG1}
\end{figure}
\end{comment}

\begin{center}
{\begin{figure}[h]
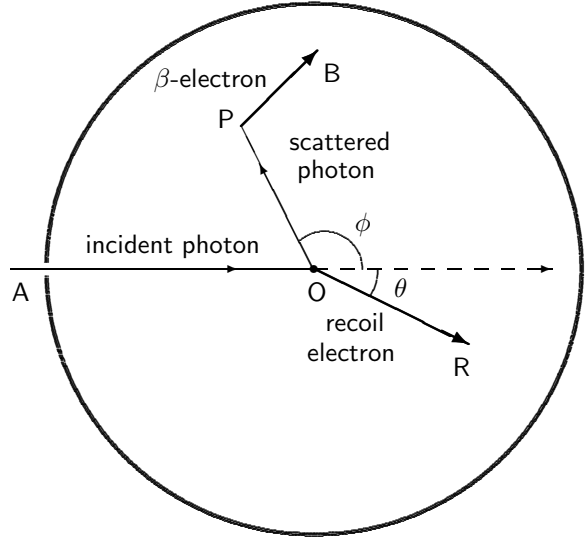


\setlength{\unitlength}{1mm}

\beginpicture

\setcoordinatesystem units <1mm,1mm>

\setplotarea x from -70 to 30, y from -37 to 32 % original

\put{\vector(1,0){30}} [B1] at -25 0 
\putrule from -10 0 to 0 0

\setdashes

\putrule from 0 0 to 30 0

\setsolid

\setlinear

\put{\vector(1,0){2.5}} [B1] at 30 0

\put{\line(-1,2){9.5}} [B1] at -3.98 0
\put{\vector(-1,2){7}} [B1] at -2.8 0

\put{\circle*{1}} [B1] at 0.5 0

{\linethickness=2pt
\circulararc 357 degrees from -35 -1 center at 0 0 
\circulararc 357 degrees from -35.2 -1 center at 0 0
\circulararc 357 degrees from -35.35 -1 center at 0 0  
}

\circulararc 132 degrees from 5 0 center at 0 0
\circulararc -30 degrees from 7 0 center at 0 0

\thicklines

\put{\vector(2,-1){20}} [B1] at 8.9 0
\put{\vector(1,1){10}} [B1] at -6 19
\put{\vector(2,-1){20}} [B1] at 8.95 0
\put{\vector(1,1){10}} [B1] at -5.95 19

\put{\footnotesize$\theta$} at 10 -2.5
\put{\footnotesize$\phi$} at 5 6

\footnotesize
\put {\sf O} at -1 -3
\put {\sf A} at -40 -3
\put {\sf P} at -13 20
\put {\sf R} at 18 -13
\put {\sf B} at 1 26

\put{\sf incident photon} at -20 3

\put{\sf scattered} at 2 17
\put{\sf photon} at 1.5 13

\put{\sf recoil} at 4 -7
\put{\sf electron} at 3.5 -11

\put{\sf $\beta$-electron} at -15 25

\endpicture
\caption{Illustrating the Compton-Simon experiment}\label{FIG}
\end{figure}
}
\end{center}

If energy and momenta are conserved in the scattering process, $\phi$
would be related to $\theta$ by {eq.}~(1) of \cite{CS1925} (reproduced
as eq.~(A-1) in Appendix \ref{R-CS}). If, however, they are
conserved only statistically, as suggested by BKS, the measured angles
$\phi_{\sf meas}$ would show a statistical spread around the $\phi$
given by eq.~(A-1) which would far more substantial than the
spread due to experimental errors. The experiment would clearly
distinguish between the statistical energy-momentum conservation
suggested in \cite{BKS1924} and exact conservation in individual
scattering events.

In {\bf II}, Compton and Simon state their numerical results, and in
{\bf III} they sum up their results. The experiment did not support
the BKS hypothesis that energy and momentum were conserved only
statistically; rather, they were `in direct support' of the view that
energy and momentum were conserved in individual electron-photon
collisions. The experiment itself consisted of \emph{recording the
paths of the recoil and the $\beta$-electrons on a cloud chamber
photograph, and measuring the angles $\theta$ and $\phi$ on the
photograph}. The wavelength $\lambda$ of the incident photon was
known.

\subsection{Von Neumann's reading of the experiment}\label{VN-READING}

Von Neumann's reading of the Compton-Simon experiment is summarized
below. According to von Neumann:

\begin{enumerate}

\nitem Energy and momenta of the scattered photon and recoil electron
were measured \emph{separately} in the experiment, and the
conservation laws were found to hold (see\ {\bf 1}).

\nitem The two measurements were \emph{not simultaneous} (the time
difference between them was $10^{-10}$ to $10^{-9}$ sec; see\ {\bf 4}).

\nitem The experimentalist could choose \emph{which measurement to
perform first} (see\ {\bf 4}).

\nitem Both measurements gave the same result for the direction of
momentum transfer (or any coordinate of it or of the point of
collision) in the scattering event (see\ {\bf 1--3}).

\end{enumerate}

Von Neumann stated that three degrees of causality or non-causality
may be discerned in nature (see\ {\bf 5}). First, both the initial and
the final state have a dispersion with respect to $\frR$, the quantity
observed.  Second ``the value of $\frR$ may have a dispersion in the
first measurement, but that {\bf immediately subsequent measurement is
constrained to give a result which agrees with that of the first}''
[quoted verbatim]. Third, neither the initial nor the final state has
a dispersion, i.e., $\frR$ is `determined causally at the outset'. He
then asserted that ``The Compton-Simon experiment shows that {\bf only
the second case is possible in a statistical theory}'' (see\ {\bf 6}).

\subsubsection{Two important points}\label{TWOPOINTS}

\begin{enumerate}

\item The phrase `immediately subsequent' in the quotation above is
clearly to be understood in the sense of item 2 -- a very small
interval, of the order of $10^{-10}$--$10^{-9}$ sec. If the time axis
has the structure of the real line $\bR$ (as is implicitly assumed in
every continuum-based theory) or of the rationals $\bQ$, there is no
$t^{\prime}$ that is `immediately after' any given instant $t$. Again,
in paragraph {\bf 8}, von Neumann's uses the word `immediately' in the
same sense: to mean \emph{not simultaneously, but after a very short
interval}, measured in nanoseconds, because the state may have changed
significantly after an interval measured in seconds.  

\item Von Neumann uses the term `measurement' for a measurement on a
single object, as well as a set of measurements on an ensemble of
identically-prepared objects. A measurement on a single object cannot
reveal whether or not its state has a dispersion.  But, when he
asserts that an `immediately subsequent measurement is constrained to
give a result which agrees with that of the first', he is clearly
referring to successive measurements on the \emph{same} object. 

\end{enumerate}

\subsection{Discrepancies between Compton and Simon's paper and von
Neumann's book}\label{DISCREPANCIES}

 Von Neumann asserted that the Compton-Simon experiment consisted of
two non-simultaneous energy-momentum measurements, and that both
measurements gave the same result. It was on this basis that he
claimed that ``The Compton-Simon experiment shows that {\bf only the
second case is possible in a statistical theory}''.  However, the only
quantities Compton and Simon measured were the two angles $\theta$ and
$\phi$ on the same photographic plate; they \emph{did not attempt} to
determine the energy-momentum of either the recoil electron or the
scattered photon; their experimental method would not permit it. The
question of choosing `which measurement to perform first' did not
arise!

Von Neumann not only misstated the aim of the Compton-Simon
experiment;\footnote{Footnote 124 on p 214 of his book states that
`\ldots The Compton-Simon experiment \emph{can be considered} [our
emphasis] as a refutation of this view' [of Bohr, Kramers and
Slater].} he also misunderstood the experiment itself, and read more
into its results than the experiment warranted. In short, he failed to
make a case for his assertion that `only the second case is possible
in a statistical theory'. \emph{He failed to make a case for collapse
of the state vector, \emph{and with it}, for his `first
intervention'}.

The state vector does collapse for an important class of measurements,
which is discussed in the following Section.

\section{When does the state vector collapse}\label{COLLAPSE}

In the following, Sections \ref{SEC-CONT-SPEC} and \ref{TWO-V-ON-M}
prepare the ground for addressing the question in Sec.~\ref{WHEN}.

\subsection{Measurement of observables with continuous
spectra}\label{SEC-CONT-SPEC}

Consider the measurement of a continuous variable in classical
physics, say one known to lie in the interval $(0,1)$. The measurement
will always return a rational number; however, the number returned may
be any rational in $(0,1)$. Typically, the measurement is repeated
many times, and the result expressed as $r\pm\Delta r$. Here $r$ is
the mean of the observations, the measurement is always approximate,
and no-one has any doubt about what the terms `measurement error' and
`approximate measurement' mean.

It is different in quantum mechanics. Since no normalizable
eigenvectors are associated with continuous spectra, von Neumann
approximated observables $A_{\sf cont}$ with continuous spectra by
operators $A_{\sf discr}$ with discrete rational spectra (with
associated eigenvectors) (\cite{VN1932}, p 222--223). However, the
spectrum of $A_{\sf discr}$ was nowhere dense in that of $A_{\sf
cont}$, and it was no longer possible to lift the notions of
approximate measurements and errors of measurement from classical
physics. On the other hand, as the spectra of the operators $A_{\sf
discr}$ were rational they could, presumably, be measured exactly.

\subsection{Two views on measurements}\label{TWO-V-ON-M}

A further ambiguity in the use of the term `measurement' (on a single
object) remains to be clarified. Consider the following quotations
from Wigner and Jauch.  In 1963, Wigner wrote (\cite{EPW1970}, pp
156--157):

\begin{quote}
`If one wants to describe the process of measurement by the equations
of quantum mechanics, one will have to analyze the interaction between
object and measuring apparatus. Let us consider a measurement from the
point of view in which the ``sharp'' states are $\sigma^{(1)},
\sigma^{(1)},\ldots$ For these states of the object the measurement
will surely yield the values $\lambda_1, \lambda_2, \ldots$
respectively. Let us further denote the initial state of the apparatus
by $a$; then, if the initial state of the system was $\sigma^{(\nu)}$,
the total system -- apparatus plus object -- will be characterized,
before they come into interaction, by $a \times \sigma^{(\nu)}$.
\emph{The interaction should not change the state of the object in
this case} and hence will lead to  
$$ a\times \sigma^{(\nu)} \rightarrow a^{(\nu)}\times
\sigma^{(\nu)}\eqno{\text{(Wigner-1)}}$$ 
\emph{The state of the object has not changed}, but the state of the
apparatus has and will depend on the original state of the object'
[emphases added].
\end{quote}

Jauch viewed it differently. This is what he wrote in 1968
(\cite{JAUCH1968}, p 163):

\begin{quote}
`\ldots the measuring device $M$, if it is to be of any use at all,
must interact somehow with the system $S$. \emph{But an interaction
always acts both ways} [emphasis added]. Not only does $S$ influence
$M$, thereby producing the desired measurable effect, but $M$ also
acts on $S$, producing an effect on $S$ with no particularly desirable
consequences. In fact, this back effect on $S$ seems to be the cause
of much of the difficulty in the interpretation of quantum mechanics.'
\end{quote}

Wigner assumes that it makes sense to talk about measurements that do
not disturb an object which is in an eigenstate of the observable
being measured. Jauch asserts that every measurement causes a
back-action upon the object -- it always disturbs the latter.  Jauch
is undoubtedly correct as far as actual measurements are concerned.
But mathematical descriptions often assume more structure than can be
revealed by laboratory experiments; the use of continuous variables
and differentiable functions are all-pervasive examples. We shall
remain with von Neumann's (idealized) notion of measurement as
clarified by Wigner, in which the back-action of the apparatus upon
the object is ignored.

\subsection{When must the state vector collapse?}\label{WHEN}

By bringing the Compton-Simon experiment into the picture, von Neumann
draws attention to the measurement of conserved quantities.  Clearly,
the measurement of an \emph{additively conserved quantity} on a
single, isolated object will necessarily cause the latter's state
vector to collapse to an eigenvector belonging to the measured
eigenvalue; if it does not, a subsequent measurement (on the same
object) may yield a different eigenvalue, \emph{which will violate the
conservation law} (\cite{SEN2010}, p 140). Collapse may therefore be
seen as the price of reconciling the superposition principle with
additive conservation laws. Since the latter are rooted in symmetry
principles of classical physics, and not in the basic principles of
quantum mechanics itself, what really remains to be established is
whether they are consistent with the superposition principle and
quantum dynamics.

However, study of the measurement problem has followed a somewhat
different trajectory, and it is useful to provide a brief historical
overview.

\section{Is the quantum measurement problem soluble or
insoluble?}\label{QMP}

\subsection{The problem}

We begin with what seems to be a reasonable definition from the point
of view of physics:

\begin{defn}[Quantum measurement problem] The problem of 
explaining the collapse of the state vector {\rm within Schr\"odinger
dynamics}, without introducing the observer's conscious ego, will be
called the \emph{quantum measurement problem {\rm{(QMP)}}.} 
\end{defn}

The physicist would like to get rid of von Neumann's first
intervention, and with it, of the observer's conscious ego; there
should be only `one Ring to bind them all',  one temporal
evolution.\footnote{To be absolutely clear, by `temporal evolution' we
mean evolution according to a Schr\"odinger equation; the hamiltonian
may or may not be time-independent -- both cases are included.}
Unfortunately, the physical problem does not define a unique
mathematical problem.

Between 1970 and 1997, Fine \cite{FINE1970}, Shimony \cite{AS1974},
Brown \cite{BROWN1986}, Busch and Shimony \cite{BS1996} and Stein
\cite{STEIN1997} offered proofs that the QMP is insoluble.  Then, in
2005, Sewell offered an explicit solution in \cite{GLS2005}.  Since a
well-defined mathematical problem cannot simultaneously be insoluble
and have a solution, these authors must have been addressing problems
that were mathematically different; and indeed they were; they
differed fundamentally on the nature of the measuring apparatus!

\subsection{Two views on the apparatus}\label{APPARATUS}

Quantum mechanics provides no characterization of measuring
instruments, and two very different schools of thought have emerged.
According to Bohr, the result of any measurement must be expressed in
classical terms.  But, according to Wigner's elaboration of von
Neumann (as already quoted in the Introduction), states of the
measuring apparatus do not have `classical description[s]'. As Shimony
wrote \cite{AS1974}:

\begin{quote}
`The insistence upon a classical description of the measuring
apparatus, not as a convenient approximation but as a matter of
principle, clearly differentiates Bohr's interpretation of quantum
mechanics from that of von Neumann and of London and Bauer.'
\end{quote}

Wigner's view was discussed by him in \cite{EPW1963} and in
\cite{EPW1964}, and in greater detail in lectures given in Princeton
\cite{EPW1976}. In these lectures he concedes a possible need for a
`large apparatus', but not that it may have a classical description
(whatever that may mean). London and Bauer's presentation of von
Neumann's measurement theory was given in \cite{LB1939}. Bohr held
steadfastly to his ideas, which he first articulated even before
quantum mechanics came into existence. A brief account is given in
Appendix \ref{BOHR-VIEWS}.
 
We shall use the term `von Neumann-Wigner measurement problem' to
denote the one in which any self-adjoint operator is measured and the
apparatus need not have a classical description; and the term `Bohr
measurement problem' to denote the one in which an additively
conserved quantity is measured with an apparatus which \emph{does}
have a classical description, as formalized in paragraph 1 of
Sec.~\ref{SEWELL}.

\subsection{The von Neumann-Wigner measurement problem is
insoluble}\label{INSOLUBLE}

In accordance with the remarks in Sec.~\ref{SEC-CONT-SPEC}, we shall
be concerned only with the \emph{measurement of observables with
discrete spectra} in what follows. Additionally, we shall assume that
the object and apparatus together constitute a \emph{closed and
isolated system}.

Let $\frHI$ and $\frHII$ be the Hilbert spaces of the object and
apparatus respectively, and $\phI\otimes\phII$ the initial state of
the object-apparatus system, where $\phI\in\frHI$ and $\phII\in
\frHII$.  In von Neumann's measurement theory, the measurement
interaction is left unspecified, but is supposed to effect,
\emph{under a Schr\"odinger equation}, the change 
\begin{equation}\label{VN-FINAL}
\phI\otimes\phII \rightarrow
\sum_{j=0}^{\infty}c_j\,\vphI_j\otimes\vphII_j
\end{equation}
where $c_j > 0$, $\vphI_j$ are eigenvectors of the operator $\AI$
being measured,\footnote{Here we are following Wigner's simplified
presentation \cite{EPW1963} of von Neumann's theory.  Textbook-level
accounts of von Neumann's original version, and of Sewell's theory
discussed in Sec.~\ref{SEWELL}, may be found in \cite{SEN2010}.} and
the right-hand side represents the final state of the object-apparatus
system. Note that this final state is a pure, and not a mixed state.

Our starting-point is a conclusion of Araki and Yanase on
limitations of the accuracy of measurements in von Neumann's theory,
following earlier work by Wigner.

\subsubsection{The results of Wigner, Araki and Yanase}\label{SEC-AY-CONC}

If the initial state of the object is $\vphI_{\mu}$, an eigenstate of
$\AI$, then (\ref{VN-FINAL}) would reduce to
$\vphI_{\mu}\otimes\phII\rightarrow \vphI_{\mu}\otimes\vphII_{\mu}$ or,
as Araki and Yanase wrote it \cite{AY1960},
\begin{equation}\label{A-Y-1}
U(t)[\vphI_{\mu}\otimes\phII] =  \vphI_{\mu}\otimes\vphII_{\mu}
\end{equation}
where $t$ represents any instant after the measurement interaction has
fully taken effect.

Following a special case established earlier by Wigner \cite{EPW1952},
Araki and Yanase proved, in 1960, that:

\begin{thm}[Wigner-Araki-Yanase] Eq.~{\rm(\ref{A-Y-1})} cannot hold
unless $\AI$ commutes with \emph{every additively conserved quantity}
$L^{\rmI}$.
\end{thm}
They then defined a quantifiable notion of the \emph{probability
$\epsilon$ of apparatus malfunction}, showed that the apparatus could
be redesigned to make $\epsilon >0$ as small as one wished (a result
also anticipated by Wigner), and defined an \emph{approximate
measurement} to mean one with a small $\epsilon$ \cite{AY1960}. With
this definition, they concluded that:

\begin{conc}[Araki-Yanase]\label{AY-CONC} Any self-adjoint operator
can always be measured approximately.
\end{conc}

In 1961 Yanase showed that, if $L^{\rmI}_z = \sigma_z^{\rmI}$
(the $z$-component of the angular momentum of the object) is
conserved, then, in any measurement of $\sigma_x^{\rmI}$ or
$\sigma_y^{\rmI}$, one must have 
$$\epsilon\geq~\displaystyle\frac{1}{8M^2},\;\; \text{where}\;\; M^2 =
\displaystyle\frac{(\xII, (L_z^{\rmII})^2\xII)}{\hbar^2}$$ 
$\xII$ being the initial state of the apparatus \cite{MMY1961}. The
superscripts {(\sf{o})} and {(\sf{a})} refer to $\frHI$ and $\frHII$
respectively, as in (\ref{VN-FINAL}).

Since the initial state of the object in the results of Wigner, Araki
and Yanase is an eigenstate of the observable being measured, their
results do not imply that the QMP has been solved. However, \emph{if
conclusion \emph{\ref{AY-CONC}} can be overturned}, it would imply
that the QMP is insoluble (under the given assumptions). 

The possibility of overturning \ref{AY-CONC} resides in the definition
of approximate measurement adopted by its authors. As pointed out in
Sec.~\ref{SEC-CONT-SPEC}, the classical notion of approximate
measurements cannot be lifted to quantum mechanics. Brown
\cite{BROWN1986} regarded von Neumann's main mathematical result in
chapter VI of his book, the equation for $\Phi(q,r)$ on page 434 of
\cite{VN1932}, as the `original insolubility proof' for exact
measurements. The efforts of Fine, Shimony, Brown and others were
accordingly directed at approximate measurements.  Shimony and Busch,
and Fine and Brown, showed that if the Araki-Yanase notion of
approximate measurements is replaced by others that may, arguably, be
closer to physical intuition, then their conclusion \ref{AY-CONC}
would no longer hold.  The assertion that the von Neumann-Wigner
measurement problem is insoluble is contingent upon this
understanding.

\subsubsection{The results of Shimony and Busch}\label{S-B} 

Araki and Yanase took the final state of the apparatus to be a vector
$\vphII_{\mu}\in\frHII$; Shimony wanted it to be a density matrix on
$\frHII$ (which would require the initial state of the apparatus to be
a density matrix as well), so that he could use the following
notion of approximate measurements: 
\begin{quote}

`If the initial state of the object is an eigenstate of the object
observable [$\AI$] with eigenvalue $\lambda_m$, then the final
statistical state of the object plus apparatus can be described as a
mixture of pure quantum states, all of which are eigenstates of the
apparatus observable [$I^{\rmI}\otimes\AII$], and the total
statistical weight in the mixture of those eigenstates associated with
the eigenvalue $\mu_m$ is close to $1$.  (It is understood that $m\neq
n$ implies both $\lambda_m\neq\lambda_n$ and $\mu_m\neq\mu_n$.) Hence,
the value of the apparatus observable at the end of the [measurement]
interaction\ldots is strongly correlated with the initial value of the
object observable \cite{AS1974}.'
\end{quote}

\noindent Note that neither the object observable nor the apparatus
observable is required to have a `classical description', and the
object observable need not be additively conserved.

We need some further notations to express Shimony's notion in
mathematical terms. 

\begin{enumerate}

\nitem Let $\frHI=\oplus_k\frHI_k\; \text{and}\; \frHII =
\oplus_k\frHII_k$, where $\frHI_k$ and $\frHI_k$ are eigen\-spaces of
$\AI$ and $\AII$ with eigenvalues $\lambda_k$ and $\mu_k$ respectively.
The index $k$ can take a finite or countable number of values. 

\nitem Let $\{\phI_{\alpha}\}$ be a countable set of normalized
vectors spanning $\frHI$, and $\{\vphII_j\}$ be the set of normalized
eigenvectors of $\AII$ on $\frHII$. 

\nitem Finally, let $P_{[\xi]}$ be the projection operator on the
Hilbert space $\frH$ which picks up the state $\xi\in\frH$.

\end{enumerate}

\noindent Then Shimony's notion of an approximate measurement can be
expressed as follows.

\begin{defn}[Shimony's definition of approximate 
measurement]\label{DEF-SHIMONY-APPROX-MEAS}
There exists a density matrix $T$ on $\frHII$ and a unitary operator
$U$ on $\frHI\otimes\frHII$ such that, for every $\phI\in\frHI_m$, 
\begin{equation}\label{AM1}
U(P_{[\phI]}\otimes T)U^{-1} =
\sum_{\alpha,n}a_{\alpha,n}P_{[\phI_{\alpha}\otimes\vphII_n]}                 
\end{equation}
where $a_{\alpha,n}\geq 0$, $\sum_{\alpha,n}a_{\alpha,n} = 1$, and
\begin{equation}\label{AM2}
\sum_{{\alpha,n}\atop{(n\neq m)}}a_{\alpha,n} \ll 1.
\end{equation}
\end{defn}

Shimony's insolubility theorem may now be formulated as follows:

\begin{thm}[Shimony's insolubility theorem]\label{SHIMONY-THM}
If $\AI$ has two or more distinct eigenvalues then, with def.\
\emph{\ref{DEF-SHIMONY-APPROX-MEAS}} of approximate measurements,
`there exist initial states of the object for which the final
statistical state of the object plus apparatus is not expressible as a
mixture of eigenstates of the apparatus observable'
\emph{\cite{AS1974}}.
\end{thm}

Unsharp observables can also be defined using positive-operator-valued
measures (POVM) instead of projection-valued measures. In 1996, Busch
and Shimony showed that Shimony's insolubility theorem holds even if
observables are defined by POVMs \cite{BS1996}. It is worth repeating
the last sentence of their abstract: `Both theorems show that the
measurement problem is not a consequence of neglecting the
ever-present imperfections of actual measurements.'

\subsubsection{The result of Fine and Brown}

In 1970, Fine gave a very general definition of measurement and
asserted that, under this definition, the QMP was insoluble
\cite{FINE1970}. His proof was incomplete. In 1986, Brown gave a very
simple proof \cite{BROWN1986}, using an extra assumption which he
called `real unitary evolution (RUE). We make the statements precise
in the following.  First, some notations.

As before, let $\AI$ be the observable on $\frHI$ being measured, and
$\AII$ on $\frHII$ the corresponding pointer observable. Let $\WI_{\sf
i}$ and $\WI_{\sf i^{\prime}}$ be $\AI$-distinct initial
states\footnote{If $A=\sum_k\lambda_kP_{[k]}$ is a self-adjoint
operator on $\frH$, the density matrices $W,\;W^{\prime}$ on $\frH$
are said to be $A$-\emph{distinct} if $\text{tr} (WP_{[k]}) \neq
\text{tr} (W^{\prime}P_{[k]})$ for some $k$.} on $\frHI$ and
$\WII_{\sf f}$ the final state on $\frHII$. Then:

\begin{defn}[Fine-Brown definition of measurement] The unitary
operator $U$ on $\frHI\otimes\frHII$ is called a $\langle \AI, \AII,
\WII_{\sf k}\rangle$-\emph{measurement} if and only if $U(\WI_{\sf
k}\otimes \WII_{\sf k})$ and $U(\WI_{\sf k^{\prime}}\otimes \WII_{\sf
k})$ are $(I^{\rmI}\otimes\AII)$-distinct whenever $\WI_{\sf k}$ and
$\WI_{\sf k^{\prime}}$ are $\AI$-distinct.\label{Q-MEAS}
\end{defn}  

It is well known that a non-idempotent density matrix on a Hilbert
space $\frH$ does not have a unique resolution into a weighted sum of
of projection operators onto one-dimensional subspaces of $\frH$; the
right-hand side of Shimony's definition (\ref{AM1}) is not the only
possible resolution of its left-hand side. Viewing this as a possible
weakness, Brown attempted to eliminate it by adding the RUE
hypothesis:

\begin{hypo}[Real unitary evolution, RUE]\label{RUE-HYPO}
Let the state of the object plus apparatus system  at $t=0$
be
\begin{equation}
W_0 = \sum\limits_n\,c_nP_{[\phi_n]},\;\text{\rm where}\;
c_n>0,\;\sum \limits_nc_n = 1
\end{equation}
and the $\phi_n$ are not necessary orthogonal. Let the system evolve
freely under $U=\exp (-\rmi Ht/\hbar)$, where $H$ is the free
hamiltonian of the system.  Then, at $t=\tau$,
\begin{equation}
W_{\tau} = \sum\limits_n c_nP_{[U\phi_n]}
\end{equation}
\end{hypo}

Using the RUE hypothesis, Brown gave a very simple proof of the
following theorem:

\begin{thm}[Brown's insolubility theorem] If the unitary operator $U$
on $\frHI\otimes\frHII$ corresponds to a $\langle \AI, \AII, \WII_{\sf
k}\rangle$-\emph{measurement} as defined by {\rm\ref{Q-MEAS}}, then
the final state of the {\sf O+A}-ensemble cannot be described as a
mixture of $I^{\rmI}\otimes\AII$-eigenstates for all initial
{\sf O}-ensembles.
\end{thm}

Although there are differences in the assumptions of Shimony and
Brown, their final results agree: no observable can be measured even
approximately, so that \emph{the collapse hypothesis becomes moot.}

%%%%%%%%%%%%%%%%%%%%%%%%%%%%%%%%%%%%%%%%%%%%%%%%%%%%%%%%%%%%%%%%%%%%%%%%
%%%%%%%%%%%%%%%%%%%%%%%%%%%%%%%%%%%%%%%%%%%%%%%%%%%%%%%%%%%%%%%%%%%%%%%%

\subsection{The Bohr measurement problem is soluble}\label{SOLUTION}

For additively conserved quantities, QMP is soluble \emph{if the
apparatus is macroscopic}.  The solution, given by Sewell in 2005
\cite{GLS2005}, is outlined below.\footnote{Sewell announced his
theory for measurements of the `first kind' as defined by Jauch
\cite{JAUCH1968}, in which a second measurement, performed immediately
after the first (Jauch gives the example of position measurements),
gives the same result as the first. The phrase `immediately after' is
used in von Neumann's sense (see the first paragraph of
Sec.~\ref{TWOPOINTS}). Sewell's solution is valid for \emph{all}
observables, but it requires a modification of the probability
interpretation; the quantity measured is not an eigenvalue, but a
diagonal matrix element of the observable in a basis in which $\Ho$ is
diagonal, and collapse of the state vector has to be interpreted
accordingly (\cite{GLS2005}; also \cite{SEN2010}, Sec.~10.5.2).
For the special case of additively conserved observables, (i)~the
second measurement need not be `immediately after' the first, (ii)~the
probability interpretation holds in its standard form (measurements
return eigenvalues), and (iii)~the state vector collapses to an
eigenvector.\label{GLS-SOLN} Sewell's original theory is at variance
with the insolubility theorems for the von Neumann-Wigner measurement
problem, and may be better suited to describe real-life measurements.}
In Sewell's theory, object plus apparatus form a closed, conservative
system. The temporal evolution of the coupled system is given by a
Schr\"odinger-von Neumann equation with an explicitly time-dependent
hamiltonian; the interaction hamiltonian acts only between $t=0$ and
$t=\tau$, where $\tau$ is small.  The macroscopic state of the
apparatus is stable under thermodynamic fluctuations in the apparatus
itself; it no longer changes after the measurement interaction has run
its course.

%%%%%%%%%%%%%%%%%%%%%%%%%%%%%%%%%%%%%%%%%%%%%%%%%%%%%%%%%%%%%%%%%%%%%% 

\subsubsection{Sewell's solution for conserved quantities}\label{SEWELL}

We shall sketch only the essentials of Sewell's theory (restricted to
the measurement of conserved quantities) in the following. Detailed
calculations may be found in chapter 10 of \cite{SEN2010}.

\paragraph*{1. The system: object and apparatus}\quad\\

\vspace{-0.5em} \noindent 
In this theory, $\frHI$ is $n$-dimensional, $n$ being small, whereas
$\frHII$ is $N$-dimensional, $N$ being of the order of Avogadro's
number; the object $\sf O$ is microscopic, whereas the apparatus {\sf
A} is macroscopic. The object plus apparatus system is denoted by {\sf
S}.

The algebras of observables on $\frHI$ and $\frHII$ are denoted by
$\cO$ and $\cA$ respectively. The quantity to be measured is
$O\in\cO$. The space $\frHI$ is spanned by its eigenvectors $\{u_i\}$,
assumed nondegenerate for simplicity.\footnote{In Sewell's original
theory, the $\{u_i\}$ are eigenvectors of $\Ho$, but not necessarily
of $O$, as $O$ is not assumed conserved. The assumption of
nondegeneracy can be dropped; see \cite{SEN2010}, pp 195--197.} Since
$O$ is conserved, the $\{u_i\}$ are also eigenvectors of $\Ho$, 
\begin{equation}\label{HO-EIGENVAL}
\Ho u_i = \epsilon_i u_i
\end{equation}
and any unit vector $\psi\in\frHI$ can be written as 
\begin{equation}\label{UNITVECTOR}
\psi = \sum\limits_{r=1}^n\,c_ru_r,\quad\mathrm{with}\quad 
\sum\limits_{r=1}^n\,|c_r|^2 = 1
\end{equation}

The algebra $\cA$ is rich in subalgebras of commuting
observables.\footnote{Von Neumann determined commuting self-adjoint
operators $\Ph, \Qh$ with discrete spectra such that $P-\Ph$ and
$Q-\Qh$ were of the order of $\hbar$ (in the sense defined by von
Neumann in \cite{VN1932}, pp~404--407, reproduced in chapter 9 of
\cite{SEN2010}). Any set $\{F(\Ph_j, \Qh_k)|1\leq j,k\leq N^{\prime},
N^{\prime}\leq N\}$ of well-defined operator functions $F$ would
constitute a commutative subset of $\cA$.} We choose from them a
finitely-generated commutative subalgebra with identity, $\cM$, of
\emph{macroscopic observables}.\footnote{The macroscopicality of these
observables is elucidated by Sewell's statistical-mechanical
description of $\cM$ in \cite{GLS2005}, following the scheme of Emch
\cite{GGE1964} and van Kampen \cite{KAMPEN}.} Let $M\in\cM$.  There
exists a set of projection operators $\{\Pi_{\alpha}|
\alpha=0,1,\ldots\nu\}$ on $\frHII$ such that
\begin{equation}\label{PI}
\sum\limits_{\alpha=0}^{\nu}\Pi_{\alpha} = \iII,\quad
\Pi_{\alpha}\Pi_{\beta} = \delta_{\alpha\beta}\Pi_{\alpha}\quad
\text{(no summation)}
\end{equation}
and 
\begin{equation}\label{STRUC-M}
M = \sum\limits_{\alpha=1}^{\nu}m_{\alpha}\Pi_{\alpha}\quad
\text{for}\quad M\in\cM
\end{equation}
The reason why this sum excludes $\alpha=0$ will appear later. Note
that $\alpha\neq\beta$ does \emph{not} imply that $m_{\alpha}\neq
m_{\beta}$.

The subspaces 
\begin{equation}
\frK_{\alpha} = \Pi_{\alpha}\frHII
\end{equation}
are the quantum analogues of classical phase cells. Each cell
represents a macroscopic state of $\cA$ which can be displayed or
registered on \emph{classical devices} -- pointers, numbers on
screens, computer printouts\ldots If the pointer `reading' is
$m_{\alpha}$, it means that the \emph{microscopic} states of $\cA$ all
lie in the subspace $\frK_{\alpha} = \Pi_{\alpha}\frHII$, where
$\text{dim}~\frK_{\alpha}\gg \text{dim}\,\frHI = n$. 

The above description of the apparatus is generic; making it specific
to the problem at hand requires some knowledge of the object-apparatus
interaction, which we shall consider next.

%%%%%%%%%%%%%%%%%%%%%%%%%%%%%%%%%%%%%%%%%%%%%%%%%%%%%%%%%%%%%%%%%%%%%%%

\paragraph*{2. The object-apparatus interaction and temporal
evolution}\quad\\ 

According to Wigner, non-specification of the object-apparatus
interaction was `the principal conceptual weakness' of von Neumann's
theory (see Wigner (\cite{EPW1970}, p 167). In Sewell's theory, by
contrast, the interaction of the microscopic object with the
\emph{macroscopic} apparatus is specified.  The total hamiltonian
$\Hs$ of the system $\sf S$ is, in an obvious notation
\begin{equation}\label{EQ-COUPLED}
\Hs = \Ho\otimes\iII + \iI\otimes\Ha + H^{\sf (int)}
\end{equation}
where the sum of the first two terms on the right in
(\ref{EQ-COUPLED}) is the free hamiltonian $\Hs^{\sf (free)}$ and the
interaction hamiltonian $H^{\sf (int)}$ is given by
\begin{equation}\label{H-INT}
H^{\sf (int)} = \chi(t)\,V,\quad\text{where}\quad \chi(t) = 
                 \left\{\begin{array}{ll}
                        1,&t \in (0,\tau)\\
                        0,&t \notin (0,\tau)
                        \end{array}
                 \right.%
\end{equation}
$V$ itself is time independent, i.e., $\Hs$ is time-independent in the
interval $(0,\tau)$. This means that the time evolution of the
system in $(0,\tau)$ can be calculated from the hamiltonian $\Hs$
using the integrated form of the Schr\"odinger-von Neumann equation
(Stone's theorem on one-parameter unitary groups, $U(t) = \exp\,(\rmi
Ht)$). The system evolves freely outside the interval $(0,\tau)$.

We assume that, at some $t<0$, object and apparatus have been prepared
independently in the states $\psi$ and $\Omega$ respectively, where
$\psi$ is pure and $\Omega$ mixed (density matrix), so that the
initial state of the object-plus-apparatus system {\sf S} at $t=0$ is
\begin{equation}\label{INITIAL-STATE}
\Phi(0) = P_{[\psi]}\otimes\Omega
\end{equation}
where $P_{[\psi]}$ is the projection operator onto the state
$\psi\in\frHI$.  Next, since $O$ is conserved, the object-apparatus
interaction $V$ should not induce transitions between different
eigenstates of $O$. Therefore, as is easily checked, $V$ must be of
the form
\begin{equation}\label{FORM-INT-H}
V = \sum\limits_{r=1}^n\,P_{[u_r]}\otimes V_r
\end{equation}
where $V_r$ are observables of $\cA$. (A restriction on the $V_r$ will
be imposed later.) Eq.~(\ref{EQ-COUPLED}) then takes the form
\begin{equation}\label{H-COMP}
\Hs = \sum\limits_{r=1}^n P_{[u_r]}\otimes K_r
\end{equation}
where
\begin{equation}\label{K-R}
K_r = \Ha + V_r + \epsilon_r\iII
\end{equation}

The temporal evolution of the system in the interval $(0,\tau)$ is
given by 
\begin{equation}\label{TEMP-EV-1}
\Phi(t) =  U^{\star}(t)\Phi(0)U(t),\quad\text{with}\quad
U(t) = \exp(\rmi\Hs t)
\end{equation}
and $\Phi(\tau)$ is defined as the left-hand limit 
\begin{equation}\label{PHI-TAU}
\Phi(\tau) = \lim\limits_{t \rightarrow \tau^-}\Phi(t)
\end{equation}
Using (\ref{H-COMP}) and (\ref{K-R}), one establishes by direct
calculation that for $t\in (0,\tau)$, $U(t)$ may be written as 
\begin{equation}\label{TEMP-EV-2}
U(t) = \sum\limits_{r=1}^n P_{[u_r]}\otimes U_r(t),\quad\text{where}
\quad U_r(t) = \exp (\rmi K_r t)
\end{equation}
and $\Phi(t)$ as
\begin{equation}\label{TEMP-EV-3}
\Phi(t) = \sum\limits_{r,s=1}^{n}[P_{[u_r]}P_{[\psi]}P_{[u_s]}]
\otimes [U^{\star}_r(t)\Omega U_s(t)]\\[1em]
\end{equation}

Define now an operator $R_{r,s}$ on $\Ho$ by
\begin{equation}\label{DEF-R} 
R_{r,s}\varphi = (u_s,\varphi)u_r\;\forall\;\phi\in\frHI
\end{equation}
Then (\ref{TEMP-EV-3}) may be rewritten as
\begin{equation}\label{FINAL-STATE}
\Phi(t) = \sum\limits_{r,s=1}^n
c_r\bar{c}_sR_{r,s}\otimes\Omega_{r,s}(t)
\end{equation}
where
\begin{equation}\label{DEF-OMEGA-RS}
\Omega_{r,s}(t) = U^{\star}_r(t)\Omega(0)U_s(t) 
\end{equation}
Since $\Omega$ has unit trace, so do the $\Omega_{r,r}$ forevery $r$.

It may also be verified that, if $V=0$, then $\Omega_{r,s}(t) =
U^{\star}(t)\Omega(0)U(t)$; the apparatus evolves independently of the
object. 

Eq.\ (\ref{FINAL-STATE}) follows from the Schr\"odinger-von Neumann
equation with the hamiltonian $\Hs$ (eqns.~(\ref{H-COMP}) and
(\ref{K-R})) and the definitions (\ref{DEF-R}) and
(\ref{DEF-OMEGA-RS}); the conservation laws are taken into account by
the definition of the $u_r$ and eqns.\ $(\ref{HO-EIGENVAL})$ and
$(\ref{FORM-INT-H})$.  It contains all the information we need.
However, to extract this information we need a tool developed by
Sewell. 

%%%%%%%%%%%%%%%%%%%%%%%%%%%%%%%%%%%%%%%%%%%%%%%%%%%%%%%%%%%%%%%%%%%%%%%%%  

\paragraph*{3. Expectation values and conditional expectations of
observables}\quad\\

\vspace{-0.5em} \noindent With the initial state of the object being a
superposition of different eigenstates of $O$, suppose that two
successive measurements of $O$ on the same object have yielded the
values $\lambda_1$ and $\lambda_2$. If $\lambda_1 = \lambda_2 =
\lambda$, one would infer that the measurement has caused the wave
packet to collapse. This inference does not require one \emph{to know
the value of $\lambda$} -- only whether or not
$\lambda_1=\lambda_2$.\footnote{This crucial fact was first noticed by
Sewell in \cite{GLS2005}.} The Schr\"odinger equation does not
determine the value of $\lambda$; it does, however determine whether
or not $\lambda_1 = \lambda_2$, as we shall soon see. We begin with
some preliminary notations and definitions.

The expectation value of the observable $O\otimes M$ at time $t$, i.e.,
in the state $\Phi(t)$, is given by
\begin{equation}\label{EXPECT-1}
E(O\otimes M) = \text{tr}\left[\Phi(t)(O\otimes M)\right]
\end{equation}
Important special cases of the general formula (\ref{EXPECT-1}) will
be denoted by the following shorthands:
\begin{equation}\label{EXPECT-2}
E(O) = E(O\otimes\iII), \quad E(M) = E(\iI\otimes M)
\end{equation}
$E(O)$ is the \emph{unconditional} time-dependent expectation value of
the observable $O$ of the object, unconditional because it uses no
information provided by the apparatus. Next, since $\Pi_{\alpha}$ is
the projection operator onto the cell $\frK_{\alpha}$, the
time-dependent probability that the macroscopic state of the apparatus
$\sf A$ is defined by the cell $\frK_{\alpha}$ is given by
\begin{equation}\label{W-ALPHA}
w_{\alpha}(t) = E(\iI\otimes\Pi_{\alpha})
\end{equation}
We single out the index $\alpha=0$ to denote the rest state $\Omega$
of the apparatus at $t=0$
\begin{equation}\label{REST-STATE}
w_{\alpha}(0) = \delta_{\alpha,0}
\end{equation}

Sewell noticed that the dependence of the result of the \emph{second}
measurement on the result of the \emph{first} measurement could be
expressed in terms of a \emph{conditional expectation}. In classical
probability theory, conditional expectations are defined as follows:

\begin{defn}[Expectation conditioned on a ${\sigma}$-subalgebra] Let 
$\mathcal{F}$\\ be a $\sigma$-algebra, $\mathcal{G}$ a
$\sigma$-subalgebra of it, and $\bf{X}$ a random variable on
$\mathcal{F}$.  The \emph{conditional expectation} of $\bf X$ given
$\mathcal{G}$ is the unique random variable
$\bf{E}(\bf{X}|\mathcal{G})$ on $\mathcal{G}$ that satisfies 
$$ E{(\bf{E}(\bf{X}|\mathcal{G})\bf{Y})} = {E}(\bf{XY}) $$
\end{defn}
\noindent Extension to noncommutative algebras may present
difficulties, but for the algebras $\cO$ and $\cM$ defined earlier,
Sewell was able to prove, by direct construction, that

\begin{thm}[Sewell] There exists a unique positive linear functional
${\bf{E}(\cdot|\cM)}:\cO\rightarrow\cM$ that preserves normalization
and satisfies
\begin{equation}\label{COND-EXPECT}
E({\bf{E}}(O|\cM)M)) = E(O\otimes M) \;\;\forall\;\;O\in\cO,\;M\in\cM 
\end{equation}
\end{thm}
${\bf{E}}(O|\cM)$ is the conditional expectation of $O$ given $\cM$ 
(`given $\cM$' meaning that the first measurement has been
completed).  Since ${\bf{E}}(O|\cM) \in \cM$, it may, according to
(\ref{STRUC-M}), be written as
\begin{equation}\label{FORM-COND-EXPECT}
{\bf{E}}(O|\cM) = \sum\limits_{\alpha}\,f_{\alpha}(O)\Pi_{\alpha}
\end{equation}
The coefficient $f_{\alpha}(O)$ of $\Pi_{\alpha}$ in
(\ref{FORM-COND-EXPECT}) is the expectation value of $O$ when the
apparatus is in the macrostate defined by the subspace
$\frK_{\alpha}\subset\frHII$. Denoting this expectation value by
$E(O|\frK_{\alpha})$, we have
\begin{equation}\label{DEF-E-O-K-A}
E(O|\frK_{\alpha}) = f_{\alpha}(O)
\end{equation}

Sewell obtained the following explicit expression for
$E(O|\frK_{\alpha})$ (for the derivation, see pp~183--184 of
\cite{SEN2010})
\begin{equation}\label{CE-OA}
E(O|\frK_{\alpha}) = \frac{E(O\otimes\Pi_{\alpha})}
{w_{\alpha}}\quad\forall\quad O\in \cO,\; w_{\alpha} \neq 0
\end{equation}

\paragraph*{4. Calculating expectation values}\quad\\

\vspace{-0.5em} \noindent We have to evaluate $E(O\otimes M)$, i.e.,
calculate the trace on the right-hand side of (\ref{EXPECT-1}). 

Define
\begin{equation}\label{DEF-F}
F_{r,s;\alpha}(t) = \text{tr}\,(\Omega_{r,s}(t)\Pi_{\alpha})
\end{equation}
The $F_{r,r;\alpha}$ satisfy 
\begin{equation}\label{FRR-BOUNDS}
0\leq F_{r,r;\alpha}\leq 1
\end{equation}
and
\begin{equation}\label{FRR-SUM}
\sum\limits_{\alpha=1}^{\nu}F_{r,r;\alpha} = 1
\end{equation}
The $F$'s also satisfy, for each $\alpha$, the inequality
\begin{equation}\label{F-INEQ}
F_{r,r;\alpha}F_{s,s;\alpha}\geq |F_{r,s;\alpha}|^2
\end{equation}

Explicit calculation yields the formulae
\begin{equation}\label{EAM}
\begin{array}{rcl}
E(O\otimes \Pi_{\alpha}) &=&
\sum\limits_{r=1}^n\,
|c_r|^2\,\lambda_rF_{r,r;\alpha}(t)\\[6mm]
E(O\otimes M) &=& \sum\limits_{r=1}^n\sum\limits_{\alpha=0}^{\nu}\,
|c_r|^2\,\lambda_r\,m_{\alpha}F_{r,r;\alpha}(t)
\end{array}%
\end{equation}
where $\lambda_r = (u_r,Ou_r)$. From the definition (\ref{W-ALPHA}) of
$w_{\alpha}$, we find, from the first equation of (\ref{EAM}),
\begin{equation}\label{W-ALPHA-2}
w_{\alpha}(t) = \sum\limits_{r=1}^n|c_r|^2F_{r,r;\alpha}(t)
\end{equation}
Recall that $w_{\alpha}(t)$ is the probability that the macroscopic
state of the apparatus $\sf A$ is defined by the cell $\frK_{\alpha}$.

The quantities $F_{r,s;\alpha}$ for $r\neq s$ do not appear in
(\ref{EAM}) or in (\ref{W-ALPHA-2}); then (\ref{DEF-F}) shows that the
$\Omega_{r,s}$ with $r\neq s$ do not contribute to the traces
$F_{r,s;\alpha}$.

%%%%%%%%%%%%%%%%%%%%%%%%%%%%%%%%%%%%%%%%%%%%%%%%%%%%%%%%%%%%%%%%%%%%%%

\paragraph*{5. Adapting the apparatus; condition for measurement}\quad\\

\vspace{-0.5em} \noindent We assume, for simplicity, that the
apparatus measures only the single observable $O$, i.e., $\cM$
consists of scalar multiples of $M$, together with the identity
$\iII$. Then (\ref{STRUC-M}) becomes the spectral decomposition of
$M$.

Since the apparatus was prepared in the initial state $\Omega$, it
follows that 
$$\alpha\neq 0 \Rightarrow \Pi_{\alpha}(\Omega\,\frHII) = 0 $$
in words, the density matrix $\Omega$ lies entirely in the cell
$\frK_0\subset\frHII$.

If the apparatus is to measure the observable $O$, the following
condition must be satisfied:
\begin{cond}[Adapting the apparatus]\label{ADAPT-APP}
To every eigenstate $u_r$ of $O$ there corresponds a unique
macroscopic state $\alpha$ of the apparatus, where $\alpha\neq 0$.
\end{cond}
After the measurement interaction has run its course, `the state of
the apparatus should reflect the state of the object'.  That is, the
following condition must be satisfied by the measurement interaction:

\begin{cond}[Condition for measurement]\label{CH-MEAS}

For $t\geq\tau$, the apparatus shall be in a unique macroscopic state
$\alpha,\; \alpha \neq 0$, corresponding to a unique $($microscopic$)$
state $u_r$ of the object.

\end{cond}

Without loss of generality, we may assume condition \ref{CH-MEAS} to
mean that there exists an invertible map $\sigma$ such that
\begin{equation}\label{SIGMA}
\alpha = \sigma(r)\quad\text{for}\quad\alpha\neq 0
\end{equation}

\noindent Conditions \ref{ADAPT-APP} and \ref{CH-MEAS} imply that
\begin{equation}\label{TR-RR}
F_{r,r;\beta}(t) = \text{tr}\,(\Omega_{r,r}\Pi_{\beta}) =
\delta_{\sigma(r),\beta}
\end{equation}
This shows that for $r\neq s$ and fixed $\alpha$, at least one of
$F_{r,r;\alpha}(t)$ and $F_{s,s;\alpha}(t)$ must vanish. Then, from
(\ref{F-INEQ}), we find that
$$ F_{r,s;\alpha}(t) = 0\;\;\text{for}\;\; r\neq s $$
which confirms that, for times $t\geq\tau$, the density matrix for the
state of the apparatus does indeed lie in a single cell
$\frK_{\alpha}$.

%%%%%%%%%%%%%%%%%%%%%%%%%%%%%%%%%%%%%%%%%%%%%%%%%%%%%%%%%%%%%%%%%%%%%%%

\paragraph*{6. Results}\quad\\

\noindent Using (\ref{TR-RR}) and setting $M=\iII$ in the second of
(\ref{EAM}), we find that
\begin{equation}\label{F-1}
E(O) = \sum\limits_{r=1}^n \lambda_r|c_r|^2
\end{equation}
Using (\ref{TR-RR}) again, we find from (\ref{CE-OA}) and
(\ref{W-ALPHA-2}) that
\begin{equation}\label{F-2}
E(O|\frK_{\alpha}) = \lambda_{\sigma^{-1}(\alpha)}
\end{equation}

\noindent Eq.~(\ref{F-1}) is the \emph{unconditional} expectation
value of $O$, as quantum mechanics would expect us to find.
Eq.~(\ref{F-2}) is the conditional expectation value of $O$, given the
state of the apparatus $\frK_{\alpha}$. It shows that the result of
the second measurement is the same as that of the first; the first
measurement has caused the wave packet of the object to collapse to
one of the eigenstates of $O$; it does not tell us \emph{which}
eigenstate, as $\alpha$, and therefore $r=\sigma^{-1}(\alpha)$, remain
unknown as long as the pointer is not `read'.\footnote{As Jauch has
pointed out \cite{JAUCH1968}, `reading the pointer' need not involve a
human observer; a computer printout would do just as well.}

If the initial state $\psi$ is an eigenstate of $O$, e.g., $\psi =
u_k$, $k$ fixed, then initial and final states (\ref{INITIAL-STATE})
and (\ref{TEMP-EV-3}) become, respectively 
\begin{equation}\label{INITIAL-SPECIAL}
\Phi(0) = P_{[u_k]}\otimes\Omega(0)
\end{equation}
and
\begin{equation}\label{FINAL-SPECIAL}
\Phi(\tau) = P_{[u_k]}\otimes\Omega_{k,k}(\tau)
\end{equation}
We see that the measurement has changed the state of the apparatus,
but not that of the object, in conformity with eq.~(Wigner-1).

This completes the solution of the Bohr measurement problem.

%%%%%%%%%%%%%%%%%%%%%%%%%%%%%%%%%%%%%%%%%%%%%%%%%%%%%%%%%%%%%%%

\paragraph*{7. Reversibility and Irreversibility}\quad\\

Measurement of $O$ on a single object prepared in the initial state
$\psi$ as given by (\ref{UNITVECTOR}) sends it, at $t=\tau$, into a
final state

\begin{equation}\label{FINAL-GENERAL}
\Phi^{[j]}(\tau) = P_{[u_j]}\otimes\Omega_{j,j}(\tau)
\end{equation}
where the value of $j$ is not predicted by quantum mechanics. The
measurement has changed only those states of $\frHII$ that lie in the
subspace $\frK_{\alpha = \sigma(j)}$. We can time-reverse this motion
by applying the operator $U(-\tau) = U^{\star}(\tau)$. For
$\Phi^{[j]}(\tau)$ as above, (\ref{TEMP-EV-2}) reduces to the single
term
\begin{equation}\label{REV-TEMP-EV}
U^{[j]}(t) =P_{[u_j]}\otimes U_j(t)
\end{equation}
so that, with $\Phi^{\sf rev}(0)$ being the limit of $\Phi(\tau)$ as
$\tau$ approaches $0$ from the right,
\begin{equation}\label{REVERSED}
\begin{array}{rcl}
\Phi^{(\sf rev)}(0) &=& U^{[j]}(\tau)\Phi(\tau)U^{[j]\star}(\tau)\\
                    &=& P_{[u_j]}\otimes
	      U_j(\tau)\Omega_{j,j}(\tau)U^{\star}_j(\tau)\\ 
              &=& P_{[u_j]} \otimes\Omega(0)

\end{array}
\end{equation}
We see that the apparatus has returned to its initial state, but the
object has remained in its final state! This establishes the
irreversibility of collapse.

\subsection{Summing-up}\label{SUMMING-UP}

With the chosen form of the time-dependent total hamiltonian, the
coupled object-apparatus system evolves differently in the
subintervals $(-\infty,0)$, $(0,\tau)$ and $(\tau, \infty)$. In the
first and the third subintervals, the evolution is given by the
integrated Schr\"odinger (or von Neumann) equation with the
\emph{free} hamiltonian $\Hs^{\sf (free)}$, while in the middle
interval $(t,\tau)$, it is given by the same equation, but with the
\emph{total} hamiltonian $\Hs^{\sf (total)}$.  The states $\Phi(0)$
and $\Phi(\tau)$ are defined, respectively, as the left-hand limits of
$\Phi(t)$ as $t\rightarrow 0^-$ and $t\rightarrow \tau^-$. The motions
in each of the open subintervals $(-\infty,0)$, $(0,\tau)$ and
$(\tau,\infty)$ are individually time-reversible, but, as
$$\Phi^{(\sf rev)}(0)\neq\Phi(0)$$
the motion in its entirety in ${(-\infty,\infty)}$ is \emph{not}
time-reversible.  At $t=\tau$, the object is in an eigenstate of $O$,
and any motion from an eigenstate, \emph{forward or backward in time},
will leave the state of the object unchanged. With the given
hamiltonian, no motion from an eigenstate $u_j$ can undo the the
collapse of the initial state. The collapse process does not require
the intervention of the observer's conscious ego.  However, the
assertion that the state vector of the object has has collapsed
neither implies nor requires knowledge of the its final state.

Our discussion has been restricted to matters of principle, strictly
`neglecting the ever-present imperfections of actual measurements'. It
has attempted to show that the Schr\"odinger equation (with an
explicitly time-dependent hamiltonian) can account for the collapse of
the state vector \emph{upon measurement of an additively-conserved
quantity}.  We would therefore suggest that -- given the fact
that quantum mechanics allows the superposition of states with
different eigenvalues of a conserved observable -- the `quantum
measurement problem' is better thought of as the problem of
establishing the consistency of additive conservation laws with
superposition principle and quantum dynamics; it does not deal with
real-life measurements at all. It may be added that this view of the
measurement problem, although presented here in a historical context,
is logically independent of the context in which it has been
presented; it stands on its own two feet.

%%%%%%%%%%%%%%%%%%%%%%%%%%%%%%%%%%%%%%%%%%%%%%%%%%%%%%%%%%%%%%%%
%%%%%%%%%%%%%%%%%%%%%%%%%%%%%%%%%%%%%%%%%%%%%%%%%%%%%%%%%%%%%%%%

\section{Epilogue: Disturbing the Universe}\label{DISTURBING}

In view of the above results, we contend that nonrelativistic quantum
mechanics is a `complete' theory. It has no flaws that need to be
corrected, either by `interpretations' or by modification of the
Schr\"odinger equation. Wave packet reduction does not `take over'
from the Schr\"odinger equation, but happens when temporal evolution
is constrained by conservation laws. As for the difficulty of
quantizing gravity, well, perhaps \emph{gravity should not be
quantized}!

Is there a case for quantizing the gravitational field? As Dyson has
pointed out \cite{DYSON2013}, the Bohr-Rosenfeld argument
\cite{BR1933} -- which Dyson cast as showing that \emph{not}
quantizing the radiation field would lead to a contradiction -- does
not extend to the gravitational field. The argument is based crucially
on the existence of both positive and negative charges, and the
nonexistence of negative masses prevents its extension to the
gravitational field.  Dyson also considered four possible experiments
to detect the graviton, and concluded that three of them were
unfeasible. He added:
\begin{quote}
`\ldots there is a fourth kind which actually exists, the Planck space
telescope, detecting polarization of the microwave background
radiation. According to Alan Guth \cite{GUTH}, the polarization of the
background radiation in an inflationary universe could provide direct
evidence of single gravitons in the primordial universe before
inflation. The results of the Planck polarization measurements are not
yet published\ldots '
\end{quote}

\noindent Dyson's article was published in 2013. To the best of the
present author's knowledge, there have been no reports of the
detection of individual gravitons so far.

In short, there may be a case for \emph{not} quantizing the
gravitational field. Could it be that dark matter is unquantized?
Maybe there are limits to the applicability of reductionism to nature?
Maybe we should seriously consider these possibilities?

To sum up, we would like to quote a little-known remark by Einstein,
made in a letter to Born dated 1 June 1948 (\cite{BE1971}, item 91, p
178):

\begin{quote}

`I should like to add that I am by no means mad about the so-called
classical system, but I do consider it necessary to do justice to the
principle of general relativity in some way or other, for its
heuristic quality is indispensible to real progress.'

\end{quote}

\noindent `Perhaps Einstein would have been less opposed to quantum
mechanics if he could be persuaded that gravity need not be
quantized' (\cite{SEN2010}. p~234).

\section*{Acknowledgements}

\addcontentsline{toc}{section}{Acknowledgements}

I would like to thank Professors Helmut Reeh, Hansj\"org Roos and
Geoffrey Sewell for correspondence on earlier versions of this
manuscript.

The title of the Prologue, `The Light is Dark Enough, is inspired by
the title `The Dark is Light Enough' of a verse play by Christopher
Fry set in the Hungarian Revolution of 1848. That of the Epilogue,
`Disturbing the Universe' is the title of of a book by Dyson, who had
`collected in this book memories extending over fifty years.'  The
phrase `One Ring to bind them all' is from Tolkien's \emph{Lord of the
Rings}.

%%%%%%%%%%%%%%%%%%%%%%%%%%%%%%%%%%%%%%%%%%%%%%%%%%%%%%%%%%%%%%%%%
%%%%%%%%%%%%%%%%%%%%%%%%%%%%%%%%%%%%%%%%%%%%%%%%%%%%%%%%%%%%%%%%%

\appendix

\section*{Appendices}

\addcontentsline{toc}{section}{Appendices}

\section{Excerpts}\label{SEC-READINGS}

Appendix~\ref{R-CS} contains three quotations from the paper by
Compton and Simon.  The experiment itself has been described in
sufficient detail in {Sec.}~\ref{CS-EXPT}. Appendix~\ref{R-VN}
contains one long quotation from pp~212--218 of von Neumann's book.
These page references are to the original Beyer translation (1955),
which had `been carefully revised by the author' [Translator's Preface,
p~vi].  The \TeX{} edition of 2018, while being much easier to read,
does depart occasionally from the Beyer translation.

\subsection{Excerpts from Compton and Simon's paper}\label{R-CS}

\numberwithin{equation}{section}

The experiment of Compton and Simon was designed to test the
hypothesis -- advanced in 1924 by Bohr, Kramers and Slater
\cite{BKS1924} (hereafter BKS) to avoid the need for radiation quanta
-- that energy and momenta were conserved only statistically, and not
in individual collisions between electrons and radiation. We provide
three quotations, labelled {\bf I}, {\bf II} and {\bf III}, from
the Compton-Simon paper \cite{CS1925}. 

Quotation {\bf I}, taken from p~290 of \cite{CS1925}, defines the
quantities (angles $\theta$ and $\phi$ of eq.~(A-1) below) they
measured, and explains how measurement of these two angles sufficed to
determine whether or not energy and momenta were conserved in the
collision (between an $x$-ray photon and an electron). 

\begin{quote}

{\bf I.} `\ldots The change of wave-length of $x$-rays when scattered
and the existence of recoil electrons associated with $x$-rays, it is
true, appear to be inconsistent with the assumption that $x$-rays
proceed in spreading waves if we retain the principle of the
conservation of momentum.\footnote{Footnote in the original: Compton
\cite{COMPTON1924}.} Bohr, Kramers and Slater,\footnote{Footnote in
the original: Bohr, Kramers and Slater \cite{BKS1924}} however, have
shown that both these phenomena and the photo-electric effect may be
reconciled with the view that radiation proceeds in spherical waves if
the conservation of energy and momentum are interpreted as statistical
principles.'

`A study of the scattering of individual $x$-ray quanta and of the
recoil electrons associated with them makes possible, however, what
seems to be a crucial test between the two views of the nature of
scattered $x$-rays.\footnote{Footnote in the original: The possibility
of such a test was suggested by W. F. G. Swann in conversation with
Bohr and one of us in November 1923.\label{SWANN}} On the idea
of radiation quanta, each scattered quantum is deflected through some
definite angle $\phi$ from its incident direction, and the electron
which deflects the quantum recoils at an angle $\theta$ given by the
relation\footnote{Footnote in the original: Debye \cite{DEBYE1923};
Compton and Hubbard \cite{CH1924}\label{DCH}}
$$
\tan {\textstyle{\frac12}}\phi = - \frac{1}{[(1+\alpha)\tan
\theta]}\eqno{\text{(A-1)}}
$$
where $\alpha = h/mc\lambda$. [Here $\lambda$ is the wavelength of the
incident $x$-ray.] Thus a particular scattered quantum can produce an
effect only in the direction determined at the moment it is scattered
and predictable from the direction in which the recoiling electron
proceeds. If, however, the the scattered $x$-rays consist of spherical
waves, they may produce effects in any direction whatever, and there
should consequently be no correlation between the direction in which
recoil electrons proceed and the directions in which the effects of
the scattered $x$-rays are observed.'

\end{quote}

Quotation {\bf II}, taken from the Abstract of their paper (p~289),
gives the main numerical data:

\begin{quote}

{\bf II.} `Of the last 850 plates, 38 show both recoil tracks and
$\beta$-tracks'.  [The electrons scattered by the deflected photons
are called `$\beta$-electrons' by the authors]\ldots in 18 cases, the
direction of scattering is within $20^{\circ}$ of that to be expected
if\ldots energy and momentum are conserved during the interaction
between the radiation and the recoil electron. This number 18 is four
times the number which would have been observed if the energy of the
scattered $x$-rays proceeded in spreading waves\ldots The chance that
this agreement with theory is accidental is about 1/250. The other 20
$\beta$-ray tracks are ascribed to stray $x$-rays and to
radioactivity\ldots'

\end{quote}

Quotation {\bf III}, the concluding paragraph of their paper (on p~299),
is reproduced in below:

\begin{quote}

{\bf III.} `These results do not appear to be reconcilable with the view
of the statistical production of recoil and photo-electrons proposed
by Bohr, Kramers and Slater.  They are, on the other hand, in direct
support of the view that \emph{energy and momentum are conserved
during the interaction between radiation and individual electrons}'
[emphasis in the original].

\end{quote}

%%%%%%%%%%%%%%%%%%%%%%%%%%%%%%%%%%%%%%%%%%%%%%%%%%%%%%%%%%%%%%%%%%%

\subsection{Excerpts from von Neumann's book}\label{R-VN}

Von Neumann's description of the Compton-Simon experiment, and the
inferences he drew from it, are spelled out in 15 paragraphs on
pp~212--218 of his book. We shall number these paragraphs {\bf 1 {\rm
to} 15} for reference. Paragraphs {\bf 1--9 {\rm and} 15} are
reproduced below in their entirety, apart from a minor omission in
paragraph~{\bf 1}, marked by an ellipsis. The omitted paragraphs {\bf
10}--{\bf 14} contain purely technical proofs that the operator $R$ of
{\bf 7} need not have a discrete spectrum.\footnote{If one takes
conservation laws into account, this result loses its significance
owing to the Wigner-Araki-Yanase theorem.}

\begin{quote}

{\bf 1.} `First, let us refer to an important experiment which Compton
and Simons\footnote{The present author has found four mentions of
`Compton and Simon' in von Neumann's book; on pp 212, 214, 214\emph{n}
and 335; in all cases `Simon' is spelt `Simons'.  The latter has been
corrected to `Simon' in the \TeX{} edition.} carried out prior to
the formulation of quantum mechanics.\footnote{Footnote 123 in the
original: [reference to Compton and Simon], \cite{CS1925}. Cf.\ the
comprehensive treatment of W.  Bothe in \cite{BOTHE1926}, Chapter 3 in
particular, \& 73.\label{123}} In this experiment, light was scattered
by electrons, and the scattering process was controlled in such a way
that the scattered light and the scattered electrons were subsequently
intercepted, and their energy and momenta measured. That is, there
ensued collisions between light quanta and electrons, and the
observer, since he measured the paths after collision, could prove
whether or not the laws of elastic collision were satisfied (\ldots
The collision calculation must naturally be carried out
relativistically.) Such a mathematical calculation was in fact
possible, because the paths before collision were known, and those
after the collision were observed.  Therefore the collision problem
was entirely determined. In order to determine the same process
mechanically, two of these four paths, and the ``central line'' of the
collision (the direction of the momentum transfer) suffices. In any
case, therefore, knowledge of 3 paths is sufficient, and the fourth
acts as a check. The experiment gave complete confirmation to the
mechanical laws of collision' [conservation of energy and momentum].

{\bf 2.} `This result can also be formulated as follows, provided we
admit the validity of the laws of collision, and regard the paths
before the collision as known. The measurement of the path of either
the light quantum or the electron after the collision suffices to
determine the position of the central line of the collision. The
Compton-Simons experiment shows that these two observations give the
same result.'

{\bf 3.} `More generally, the experiment shows that the same physical
quantity (namely, any coordinate of the place of collision or of the
direction of the central line) is measured in two different ways
(by capture of the light quantum and of the electron), and the result
is always the same.'

{\bf 4.} `These two measurements do not occur entirely simultaneously.
The light quantum and the electron do not arrive at once, and by
suitable arrangement of the measuring apparatus either process may be
observed first. The time difference is usually about $10^{-9}$ to
$10^{-10}$ seconds [the time it takes a photon to traverse three to 30
cms]. We call the first measurement $\sfM_1$ and the second
measurement $\sfM_2$. $\frR$ is the quantity measured. We then have
the following situation. Although the arrangement is of such a type
that, prior to the measurement, we can make only statistical
statements concerning $\frR$, i.e., regarding $\sfM_1, \sfM_2$ (see
the reference in [foot]note {123}), the statistical correlation
between $\sfM_1$ and $\sfM_2$ is perfectly sharp (causal); the $\frR$
value of $\sfM_1$ is certainly equal to that of $\sfM_2$. Before the
measurements $\sfM_1, \sfM_2$, therefore, both results are completely
undetermined; after $\sfM_1$ has been performed (but not $\sfM_2$), the
result of $\sfM_2$ is already determined causally and uniquely.'

{\bf 5.} `We can formulate the principle that is involved as follows:
by nature, three degrees of causality or non-causality may be
distinguished.  First, the $\frR$ value could be entirely statistical,
i.e., the result of a measurement would be predicted only
statistically; and if a second measurement were then taken immediately
after the first one, this would also have a dispersion, without regard
to the value found initially -- for example, its dispersion might be
equal to the original one.\footnote{Footnote 124 in the original: A
statistical theory of elementary processes was erected by Bohr,
Kramers and Slater on these basic concepts. Cf.\ \cite{BKS1924} and
references cited in [foot]note 123. The Compton-Simons experiment can
be considered as a refutation of this view.\label{FOOTNOTE-CS}}
Second, it is conceivable that the value of $\frR$ may have a
dispersion in the first measurement, but that immediately subsequent
measurement is constrained to give a result which agrees with that of
the first.  Third, $\frR$ could be determined causally at the outset.'

{\bf 6.} `The Compton-Simons experiment shows that only the second case
is possible in a statistical theory. Therefore, if the system is
initially found in a state in which the value of $\frR$ cannot be
predicted with certainty, then this state is transformed by a
measurement $\sfM$ of $\frR$ (in the example above, $\sfM_1$) into
another state: namely, into one in which the value of $\frR$ is
uniquely determined. Moreover, the new state, in which $\sfM$ places
the system, depends not only on the arrangement of $\sfM$, but also on
the result of the measurement of $\sfM$ (which could not be predicted
causally in the original state) -- because the value of $\frR$ in this
new state must actually be equal to this $\sfM$-result.'

{\bf 7.} `Now let $\frR$ be a quantity whose operator $R$ has a pure
discrete spectrum $\lambda_1, \lambda_2, \ldots$ with the respective
eigenfunctions $\phi_1, \phi_2, \ldots$ which then form a complete
orthonormal set. In addition, let each eigenvalue be simple (i.e., of
multiplicity 1, cf. II.6.), i.e., $\lambda_{\mu} \neq \lambda_{\nu}$
for $\mu \neq \nu$. Let us assume that we have measured $\frR$ and
found a value $\lambda^{\star}$. What is the state of the system after
the measurement?'

{\bf 8.} `By virtue of the foregoing discussion, this state must be
such, that a new measurement of $\frR$ gives the result
$\lambda^{\star}$ with certainty (of course, this measurement must be
made immediately, because after $\tau$ seconds, $\phi$ has changed to
$$ \mathrm{e}^{-\frac{2\pi\mathrm{i}}{h}\tau\cdot
H}\phi\eqno{\text{(A-2)}}$$
Cf. III.2., $H$ is the energy operator.)'

{\bf 9.} `This question, as to when the measurement of $\frR$ in the
state $\phi$ gives the value $\lambda^{\star}$ with certainty, we
shall now answer in general, without limiting assumptions on the
operator $R$.'

{\bf 10.--14.} [Omitted]

{\bf 15.} `We have then answered the question as to what happens in the
measurement of a quantity $\frR$, under the above assumptions for its
operator $R$. To be sure, the ``how'' remains unexplained for the
present. This discontinuous transition from $\psi$ into one of the
states $\phi_1, \phi_2, \ldots$ (which are independent of $\psi$,
because $\psi$ enters only into the respective probabilities $P_n =
|(\psi, \phi_n)|^2, n = 1, 2, \ldots$ of this jump) is certainly not
of the type described by the time independent Schr\"odinger equation.
This latter always results in a continuous change of $\psi$, in which
the final result is uniquely determined and is dependent on $\psi$
(cf.\ the discussion in III.2.). We shall attempt to bridge this chasm
later (cf,\ VI.).'\footnote{Footnote 125 in the original: That these
jumps are related to the ``quantum jumps'' -- concept of the older
Bohr quantum theory was recognized by Jordan \cite{JORDAN1927}.}

\end{quote}

\section{Bohr's views}\label{BOHR-VIEWS}

We now turn to Bohr's views. In 1923, \emph{two years before quantum
mechanics began to be formulated}, he wrote (see p 196 of
\cite{AP1991}):

\begin{quote}
`Every description of natural processes must be based on ideas which
have been introduced and defined by classical theory' (\cite{CW1976},
vol.\ 3, p\ 458).
\end{quote}

In 1949, in his essay in Einstein's 70th birthday volume, Bohr wrote:

\begin{quote}

`For this purpose, it is decisive to recognize that, \emph{however far
the phenomena transcend the scope of classical physical explanation,
the account of all evidence must be based on classical terms}'
(\cite{NB1949}, p~209)  [emphasis in the original].

\end{quote}

\noindent His thinking on the subject had been remarkably consistent
in the quarter-century from 1923.

In his 1927 Como lecture, Bohr stated: `Our interpretation of the
experimental material rests essentially on the classical concepts'
\cite{NB1928}. Pais wrote (\cite{AP1991}, p 315):

\begin{quote}

`Bohr's lecture at Como did not bring down the house.'

L\'eon Rosenfeld who attended the Volta conference later said: `There
was a characteristic remark by Wigner after the Como lecture, ``This
lecture will not induce any one of us to change his own meaning
[opinion] about quantum mechanics''. ' 

\end{quote}

\noindent That was five years before the appearance of von Neumann's
book!

%%%%%%%%%%%%%%%%%%%%%%%%%%%%%%%%%%%%%%%%%%%%%%%%%%%%%%%%%%%%%%%

%%%%%%%%%%%%%%%%%%%%%%%%%%%%%%%%%%%%%%%%%%%%%%%%%%%%%%%%%%%%%%%

\section{Unanswered questions}\label{UQ}

How to explain von Neumann's misreading of the Compton-Simon paper?
We conjecture that von Neumann \emph{did not have a printed copy of
the Compton-Simon paper} in front of him when he wrote his account,
but relied on his phenomenal memory.\footnote{Regarding von Neumann's
memory, see \cite{NM1992}, index item `von Neumann, John, memory
power' on p 404, and the Wikipedia article on `John von Neumann',
\cite{WIKIPEDIA}, esp.\ the comments of Herman Goldstine in Sec.~9.2,
`Eidectic memory'.} That could explain why `Simon' is misspelt
`Simons' in four places in his book (pp 212, 214, 214\emph{n} and
335).\footnote{These misprints were corrected in the second (\TeX)
edition of von Neumann's book; the editor did not see any significance
in them.} There may be hints in von Neumann's discussions with Bohr at
the Institute in Princeton (see \cite{AP1991}, p 435, paragraph 3),
but the publication by Bohr referred to by Pais is not available to
me. 

In the last paragraph of chapter 10 of his popular book
\emph{Entanglement}, Aczel writes \cite{ACZEL2003}:

\begin{quote}
`When von Neumann's seminal book appeared in English, Wigner told Abner
Shimony: ``I have learned much about quantum theory from Johnny, but
the material in his Chapter Six Johnny learnt all from me''.'
\end{quote}

I wrote to Professor Shimony about this quotation. I have mislaid his
answer, but it was essentially as follows.

\begin{quote}

It is difficult to recall with certainty events that took place over
fifty years ago, but yes, Wigner did say something of the sort. I
would only question the use of the word `all'.

\end{quote}

Chapter VI of von Neumann's book is on `The Measuring Process'. The
remarks on the Compton-Simon experiment and the conclusions to be
drawn from it are in Chapter III, on `The Quantum Statistics'. Wigner
would have noticed von Neumann's misinterpretation of the
Compton-Simon experiment in no time, which gives credence to
Rechenberg's assertion, which I learned from Helmut Reeh  quite
recently. Reeh wrote:

\begin{quote}

`A historian of science, H. Rechenberg, once told me that von Neumann's
book had been written over two different periods of time.  Rechenberg
intended to investigate that but died before doing so.' 

\end{quote}

Wigner remained a firm believer in (i)~the strictly quantum character
of the measuring apparatus, and (ii)~the essential role of the
observer's consciousness in quantum measurement theory
\cite{EPW1963,EPW1964}. To the present author, the big question is, if
Johnny learnt some, or all, of his Chapter VI from Wigner, wasn't
Wigner appraised of Chapter III? The error in it would not have
escaped the physicist Wigner!

Perhaps Rechenberg was right, and Wigner came into the picture only
during the writing of the second part of the book? But such questions
are best left to historians of physics to address.

%%%%%%%%%%%%%%%%%%%%%%%%%%%%%%%%%%%%%%%%%%%%%%%%%%%%%%%%%%%%%%%%%%%%%
\small

\section*{\quad}

\end{document}